\documentclass{pasj00}
\begin{document}
\SetRunningHead{Deguchi \& Koike}{Meter-wave Luminous Objects toward Monoceros}
\Received{2008/03/28}
\Accepted{2008/09/11; ver. 3.1 Sept. 8, 2008}

\title{Clustering of Meter-wave Luminous Objects toward Monoceros}

\author{Shuji \textsc{Deguchi},  and Kazutaka \textsc{Koike}}
\affil{Nobeyama Radio Observatory, National Astronomical Observatory, \\
and Department of Astronomical Science, The Graduate University for Advanced Studies, \\
Minamimaki, Minamisaku, Nagano 384-1305 \\}
\author{[PASJ (Letter) 60 No. 6 (Dec. 25, 2008 issue) in press] }
\KeyWords{galaxies: clusters: general --- radio continuum: galaxies--- X-rays: galaxies: clusters
}  

\maketitle

\begin{abstract}
A distribution of the meter-wave luminous objects, which are bright at frequency 74 MHz 
(a wavelength of 4 m)  but not detectable at 1.4 GHz (21 cm) in the VLA surveys, 
shows a notable concentration in a scale of a few degrees
toward Monoceros [($l$, $b$)=($225^{\circ}$, $4^{\circ}$)].  
We argue that it is a part of giant radio relics associated 
with a nearby cluster of galaxies with $cz\sim 2400$ km s$^{-1}$ centered on the spiral galaxy NGC 2377.  
The angular separation of these objects from the clustering center is consistent with 
the separation of distant relics to the cluster center if scaled by distance.  
This fact implies that the concentrations of meter-wave luminous objects 
can be used as a tracer of the structure of the Local Supercluster and it's vicinity.
\end{abstract}

\section{Introduction\label{sec:Intro}}
The nature of very low frequency radio sources are not well investigated.
Especially, objects detected by the VLA Low-frequency Sky Survey (VLSS; \cite{coh07}) at 74
MHz,  but without the 1.4 GHz NRAO VLA Sky Survey  (NVSS; \cite{con98}) association, 
are quite interesting, because none of them have clear optical identifications. 
Hereafter, we call these objects as Meter-wave Luminous Objects (MLOs). 
They are considered to have extremely steep radio spectra (with spectral index $\alpha \gtrsim 2 $, 
where $I_{\nu}\sim \nu^{-\alpha}$), so that they are not detectable at 1.4 GHz. 
They are supposed to be remnants of merging galaxies \citep{con05}, 
radio halos and relics \citep{coh06}, or ultra-steep-spectrum radio sources (USSRs; \cite{gop05}). 
A sample of MLOs may also involve Galactic objects
(such as pulsars, M-dwarfs, and possibly radio-loud extra-solar planets). 
Because the nature of MLOs is not well known, it is important to establish the association 
of MLOs to the known objects.

Radio relics are diffuse radio sources which are located at the peripheral regions 
of clusters of galaxies [often called as "gisht"; see a review by \citet{kem04} and \citet{gio04}]. 
To date, approximately 30  radio relics have been mapped at 1.4 GHz/325 MHz with VLA or at 610 MHz with GMRT. 
They are known to have very steep spectral indices ($\alpha \gtrsim 1.5$), 
and are usually located at peripheral regions of clusters of galaxies without any
clear optical counterparts. They are extended by 0.1--1 Mpc in scale and one or two (double) relics are often
associated with a cluster. Their radio emissions are polarized by 10--30 percents ($B= 0.1\sim 1 \mu$ Gauss
was suggested). They are quite different from radio halos of clusters of galaxies, which are normally
centered to the dense part of a cluster and are often accompanied by X-ray halos. They are also very different
from ultra-steep spectrum radio sources (USSRS), which are relatively compact and are identified to
optical galaxies \citep{gop05}. The origin of radio relics is believed to be dying synchrotron
emission produced by shocks between merging clusters of galaxies (e.g., \cite{hoe07}). 
Radio relics share the same properties as MLOs noted above, 
although the observed frequencies concerned are slightly different. 

We identified about 400  MLOs in the VLSS surveyed region 
by comparing the VLSS and NVSS catalogs, and found 
that the sample exhibits several strong concentrations in a-few-degree scales in the sky.
In this paper, we argue that the cluster appearance of the MLOs toward Monoceros
is consistent with what has been seen in higher frequency observations
of radio relics in distant clusters of galaxies. 
The clusters of galaxies with radio relics, which were studied at higher frequencies 
in the past, are relatively distant
($z \gtrsim  0.05$), because of the smaller field of view of the telescopes at the concerned frequencies
(about a degree).  In contrast, the scale of newly found clustering of sources  at 74 MHz in this paper goes up to
about 10 degrees in diameter. The relics spreading in such large angular scales 
have never been studied at higher frequencies. Therefore, we present the investigation of
 the Monoceros clustering of MLOs in this paper, and compare the properties with those of 
known relics in clusters. We conclude that the clustering of 74 MHz sources toward Monoceros
is interpreted as aged radio relics associated with the relatively close cluster of galaxies.

\begin{figure*}
\begin{center}
\FigureFile(150mm,90mm){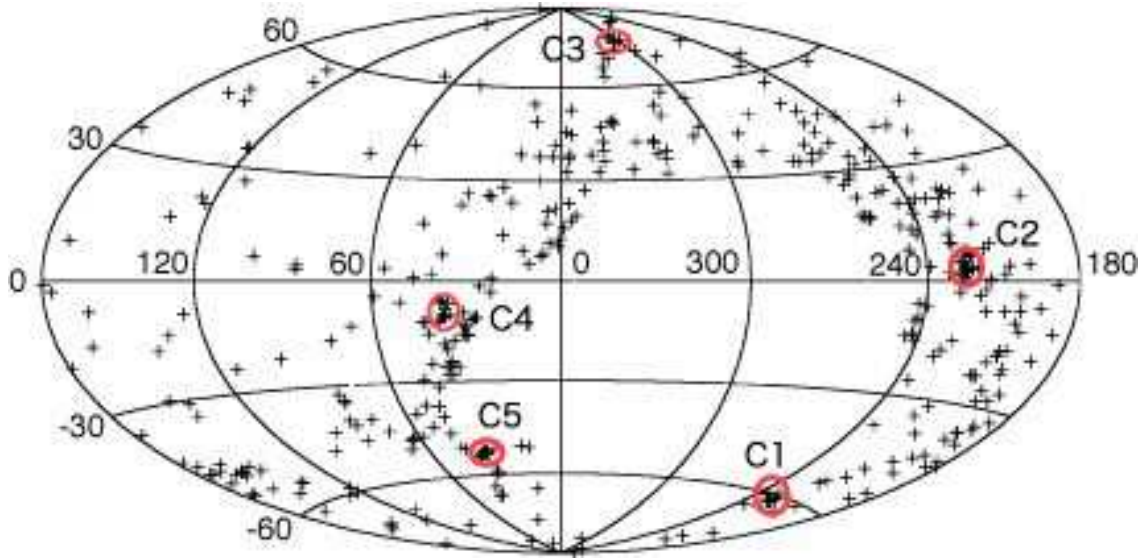}
\end{center}
\caption{Distribution of the meter-wave luminous objects in the Galactic coordinates
in the Hammer-Aitoff projection.  The cross indicates the position of MLO.
The possible concentrations of MLOs are marked by ellipses C1 -- C5. 
\label{fig:1}
}
\end{figure*}

\section{Sampling of meter-wave luminous objects from the VLSS catalog\label{sec:Obs}}
The VLSS catalog contains about 68000 radio sources,
which were detected with VLA at 74 MHz 
 with 358 $14^{\circ}\times 14^{\circ}$ continuum images (Stokes I) \citep{coh07}. 
It covers the entire sky north of $-30^{\circ}$ declination.
The flux density limit is approximately 0.7 Jy. The positional accuracy of
the source is about 20$''$ (95 \% significance; see figure 3 of \cite{coh07}). 
This is the most complete,  most wide-coverage  catalog 
at low frequencies available at present.
From this catalog, we extracted the radio sources without higher frequency detections (within a 80$''$ separation).
For such a purpose, we used the NVSS catalog,
which were also made with VLA at 1.4 GHz \citep{con98}. It covers almost
the same area of the sky observed by VLSS, and spatial positions of
the VLSS sources were aligned using the NVSS catalog positions identifying the same objects.
The NVSS catalog contains about 2 million discrete sources and the detection limit is 
about 2.5 mJy. The NVSS images have 45$''$ angular resolution (FWHM),
which is compared with the VLSS image resolution (80$''$).

Because the sensitivity of the NVSS survey exceeds that of the VLSS 
by a factor of more than 100,  the VLSS objects without NVSS detections 
are in general bright (a few Jy) at 74 MHz (4 m), but extremely faint
at 1.4 GHz (21 cm), suggesting that they have very high spectral indices ($\alpha\geq 2$).  
Therefore, we call these radio sources 
as Meter-wave Luminous Objects (MLOs).
We first selected 499 such objects from the VLSS and NVSS catalogs. However, 
we noticed that some of the strong sources without the NVSS counterparts  
in this primary list of MLOs are likely to be false (or time variable sources).   
It has been known that radio observations at very low frequencies 
are considerably influenced by ionospheric transmission, and by the large side-lobe interference effect 
of the strongest sources, such as Cyg A, Virgo A and the Galactic Center.
The positional (phase) errors due to variations of ionospheric transmission were corrected
in the VLSS catalog,  using the NVSS source positions for the phase calibration \citep{coh07}. 
However, the sidelobe interference effect does not seem to be completely removed.
\citet{coh07} claimed to remove all false detections due to the Gaussian noise
by introducing the 5 $\sigma$ peak brightness criterion,
 and further false detections due to sidelobe interferences by applying
{\it the 6 $\sigma$ peak brightness criterion} in the vicinity of sources with high peak brightness 
within a radius of
\begin{equation}
\theta _r = (1^{\circ}) \  \sqrt {(I_p \ / 60 \ {\rm Jy \ beam}^{-1})},
\end{equation}        
where $I_p$ is the peak flux density per beam of a strong source (above 12 Jy beam$^{-1}$).
Even if this procedure removed a considerable number of false detections,
we think that it was not enough. The VLSS catalog still contains
5 objects above 50 Jy and 20 objects above 5 Jy, 
which have no counterparts in the NVSS catalog. 
These bright objects are likely false detections, because they are located very near to the bright interfering sources
 (we can visually inspect their contour maps available 
in the VLSS Postage Stamp Server\footnote{available at http://www.cv.nrao.edu/4mass/VLSSpostage.shtml});
otherwise, they have much lower flux densities or time-variable flux densities.

In order to improve the reliability of the MLO sample,
we also applied the equation (1) for the selection criteria
and removed all the objects within   $\theta _r$ of the strong sources above
12 Jy beam$^{-1}$.  Thus, we selected 416 MLOs from the VLSS and NVSS catalogs,
and used this MLO sample for the later analysis.  

\begin{table*}
  \caption{MLO Cluster Candidates.}\label{tab:1}
 \begin{center}
  \begin{tabular}{cccccccl}
  \hline\hline
  No. & assignment &$l$ & $b$ & $r$ & number & density  & comment \\
       & & ($\circ$) & ($\circ$) & ($\circ$) & ($n[S/N>6]$) & (deg$^{-2})$ & \\
\hline
C1 & 0300$-$3130 & 229.1 & $-61.4$ & 3 & 12 (7) & 0.42 & void near Fornax supercluster \\
C2 & 0728$-$0900 & 225.1 & 3.9 & 4 & 20 (7) & 0.40 &Monoceros, near the north pole of SGP \\
C3 & 1254$+$1315 & 305.3 & 76.1 & 1.5 & 5 (2) & 0.71 & near Virgo A \\ 
C4 & 1933$-$0022 & 37.4  & $-9.4$ & 3.5 & 9 (4) & 0.23 & sparse, near the south pole of SGP  \\
C5 & 2212$-$2100 & 33.7 & $-53.2$ & 3 & 10 (0) & 0.35 & false feature ? \\
\hline
  \end{tabular}
  \end{center}
\end{table*} 

Figure 1 shows the distribution of MLOs in the Galactic coordinates.
The large vacant area around $l=330^{\circ}$ and $b=0^{\circ}$ is the sky
area not accessible from the VLA site ($\delta <-30^{\circ}$). 
The MLOs are clustered at several locations
which are indicated by circles C1 -- C5. Table 1 summarizes the center coordinates, 
 radius,  the number of MLOs, and average surface number density of these concentrations;
between the parentheses in column 6, the number of MLO with a high signal-to-noise ratio above 6 is shown.
 For the choice of concentration features, we use rather a rough criterion
 in this paper, i.e., more than 5 MLOs are found within a few degree radius.  
 Because the average surface number density of the MLOs between $-30^{\circ} <\delta <0^{\circ}$ 
 is approximately 0.027 per square degree,
 the density of MLO for selected clusters is more than 10 times  
 the average except the cluster C4 (which is only about 8 times the average). 
However, we noticed that the contour map toward concentration C5 (available in the VLSS postage stamp server;
 e.g., see the $3^{\circ}\times 3^{\circ}$ 25$''$ resolution contour map) exhibits a narrow ripple in source distribution;  
 4 of the 7 members near the center of concentration C5 are aligned on a straight line 
coming from 2214.4$-$1701 (about 3$^{\circ}$ away),
suggesting that they are on an interference feature.  
Therefore, a presence of the concentration C5 is questionable. 
These rippling structures appear due to insufficient UV coverage 
and incomplete deconvolution (CLEAN) procedure in the VLA observations. 

In the present MLO sample,  269 sources out of 416 MLOs are  in the southern hemisphere 
(occupying only 1/3 of the observed sky area).  
In the original VLSS catalog, we found that the $\sim 62000$ sources with 
high signal-to-noise ratio ($>6\sigma$ in the integrated intensity $S_i$) show a distribution 
with good north-south symmetry, but the  $\sim 860$ VLSS sources with low signal-to-noise ratio
($<5.5 \sigma$) show a marked  asymmetry  biased toward the southern hemisphere 
(a factor of 2--3 in density).
In fact, the south-bias tends to appear more  to the higher noise-level group (with high flux densities $S_i>1$ Jy)  
than does to the lower noise-level  group (with lower flux densities $S_i<1$ Jy) 
in the same low S/N subsample. 
Therefore, we suspect that the apparent enhancement of the MLO density
in the southern hemisphere is a problem of the VLSS detection thresholds 
in the high noise-level areas, which appear preferentially in the southern hemisphere (partially
due to insufficient UV coverage and more deconvolution errors, 
and partially due to no overlapped mapping areas near the south boundary of the sky coverage
; see section 5.2 of \cite{coh07}). 
 A substantial fraction of the low S/N southern MLOs could be false sources.
The credibility of individual sources and clusters must be assessed carefully.
If we remove the objects with low signal-to-noise ratios below 6,
the south bias of the MLO distribution is slightly mitigated, 
though the total number of the remaining MLOs is decreased to 132  
(in which 76 are in the southern hemisphere; the subsample still cannot be free from this bias).
If we apply the 6$\sigma$ criterion for the MLO selection, there still remain
7 objects in the concentration C2,\footnote
{If all 44 MLOs found within a $30^{\circ}\times 30^{\circ}$  area toward concentration C2
are randomly distributed, the probability of more than 16 objects falling in the circle of $r=4^{\circ}$
is less than 10$^{-9}$,  while it is about 1 percent for more than 7 objects being found in the same circle. 
If we take the  6$\sigma$ criterion for source selection,  the probability of 7 high $S/N$ sources falling into
the circle of 4$^{\circ}$ radius is $2.2\times 10^{-6}$ when total 13 high $S/N$ sources 
within a $30^{\circ}\times 30^{\circ}$ area in this direction are randomly distributed.
Therefore, we conclude in both cases that the concentration of C2 is not a result of statistical fluctuations.} 
though statistical significance of the other clusters except C1 and C2 is lost.
Therefore, in this paper, we only discuss  on the largest and strongest concentration C2.
The noise level is moderately low toward C2
(close to the average of the $\sim 64000$ VLSS sources below 5 Jy ), but not toward the other clusters.
Our visual inspections of the contour maps were well made and all of the sources in C2 listed in table 2 
seem to be real. 

We also tried to identify the optical/infrared counterparts for the MLOs using  
 Digitized Sky Survey (DSS),  Sloan Digital Sky Survey (SDSS), and Two-micron All Sky Survey (2MASS). 
 However, we found that none of MLOs have evident optical/infrared galaxy counterparts within 20$''$. 
 A few sources are identified with pulsars within a 5$''$ positional separation from the VLSS positions. 
 A few sources have relatively bright stellar candidates. However,
 the number of sources with a stellar candidate within 10$''$ (64\% confidence level) 
 increases with the limiting magnitude of candidate stars, and the number also decreases with $|b|$, 
 suggesting that  most of them are chance coincidence of stars along the line of sight. 
 From these facts, we deduce that only a few Galactic stars emitting 74 MHz continuum 
 are involved in the present MLO sample. From these studies,  
we believe that most of MLOs in the present sample are extragalactic objects,
such as radio relics and halos. 

\begin{table}
  \caption{Clustered sources toward Monoceros.}\label{tab:2}
 \begin{center}
  \begin{tabular}{ccccc}
  \hline\hline
  VLSS name & $l$ & $b$ & F.D.   & $\alpha _L^{*}$ \\  
       & ($\circ$) & ($\circ$) & (Jy)  & \\ 
\hline
0718.0$-$0539 & 221.004 & 3.300 & 1.28 & 2.12 \\ 
0718.3$-$1120 & 226.085 & 0.720 & 0.88 & 1.99 \\ 
0720.5$-$0757 & 223.343 & 2.792 & 1.04 & 2.05 \\ 
0721.6$-$1002 & 225.318 & 2.052 & 1.12 & 2.08 \\ 
0722.1$-$0746 & 223.354 & 3.218 & 1.25 & 2.11 \\ 
0723.4$-$1229 & 227.671 & 1.275 & 0.88 & 1.99 \\ 
0723.4$-$0904 & 224.667 & 2.889 & 2.36 & 2.33 \\ 
0724.2$-$1019 & 225.862 & 2.485 & 1.15 & 2.09 \\ 
0727.0$-$1009 & 226.048 & 3.163 & 1.60 & 2.20 \\ 
0727.5$-$1158 & 227.692 & 2.398 & 1.25 & 2.11 \\ 
0728.6$-$0921 & 225.516 & 3.886 & 1.14 & 2.08 \\ 
0732.1$-$0604 & 223.030 & 6.219 & 0.93 & 2.01 \\ 
0732.2$-$0620 & 223.291 & 6.110 & 0.90 & 2.00 \\ 
0733.1$-$0541 & 222.798 & 6.604 & 1.07 & 2.06 \\ 
0734.0$-$0949 & 226.577 & 4.845 & 0.93 & 2.01 \\ 
0734.6$-$1046 & 227.472 & 4.516 & 1.00 & 2.04 \\ 
0735.1$-$0628 & 223.734 & 6.667 & 1.08 & 2.06 \\ 
0737.0$-$0902 & 226.236 & 5.855 & 1.07 & 2.06 \\ 
0740.2$-$0757 & 225.658 & 7.065 & 1.29 & 2.12 \\ 
0742.2$-$0744 & 225.702 & 7.614 & 0.94 & 2.02 \\ 
   \hline
\multicolumn{5}{l}{$^{*}$  Lower limit of the spectral index calculated} \\ 
\multicolumn{5}{l}{from the sensitivity limit in NVSS of 2.5 mJy. }
  \end{tabular}
  \end{center}
\end{table} 


\begin{figure*}
\begin{center}
\vspace{-2.0cm}
\FigureFile(150mm,100mm){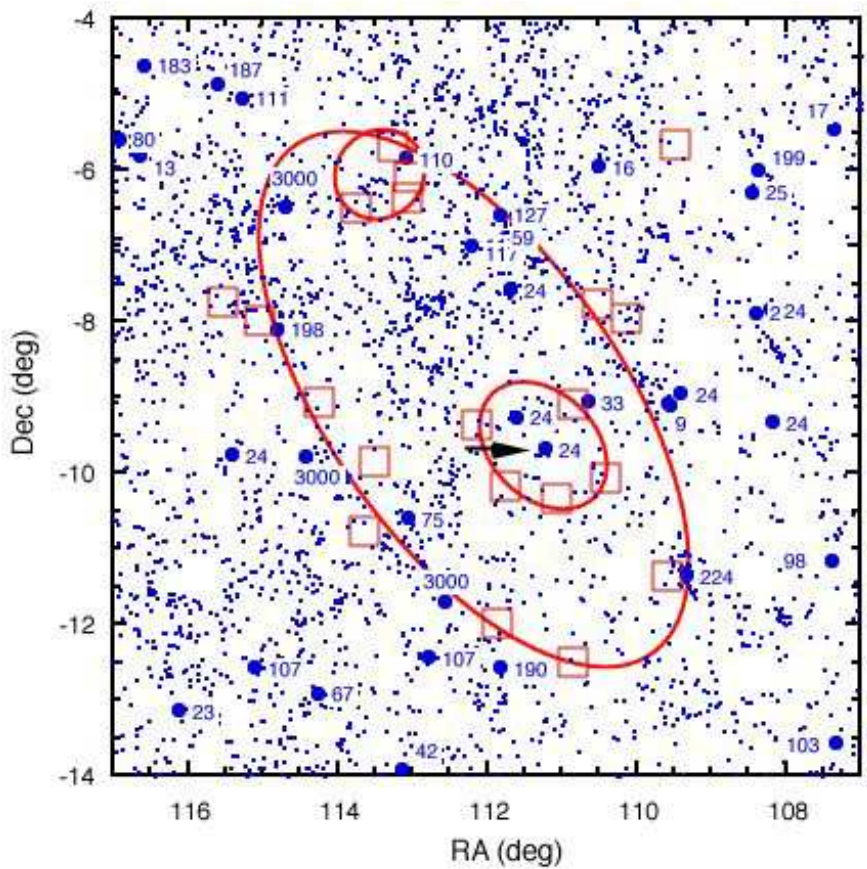}

\caption{Distribution of the meter-wave luminous objects (red squares) toward Monoceros 
in the equatorial coordinates (J2000). 
Blue  dots  are the NED galaxies identified in the optical/infrared wavelengths,  and 
blue filled circles the galaxies with known redshifts (the value of $cz$ is written in blue  in unit of 100 km s$^{-1}$,
but "3000" indicates $cz>300000$ km s$^{-1}$).
The red ellipses are possible shell structures (see in the text). The position of
NGC 2377 {\bf (=3C178)}, is indicated by the arrow.
\label{fig:2}
}
\end{center}
\end{figure*}

\section{Monoceros concentration of MLOs.\label{sec:mon}}
As mentioned in the previous section, some of the concentrations listed in table 1 might be
somewhat uncertain. However,
C2,  the largest concentration (20 sources) toward Monoceros spreading in about 9 degree
(see figure 2),  is real.
Table 2 lists their positions, flux densities in the VLSS catalog, and  the lower limit of the spectral index 
assuming the sensitivity limit of 2.5 mJy in the NVSS survey.
There is no nearby strong interference radio source (except 3C178 [=NGC 2377] with 18.7 Jy at 74 MHz). 
Because it is roughly toward the Galactic anti-center direction ($l=225^{\circ}$),  
diffuse synchrotron emission in the Galactic plane is expected 
to marginally contaminate  the VLSS sample. However,
the close location of this concentration to the Galactic plane ($b=4^{\circ}$)
makes optical identifications to galaxies somewhat difficult.
In 1990s, \citet{sai90} found that a large number of galaxies were identifiable
in the Palomer Sky Survey plates toward this direction. \citet{yam93} measured the redshift of 
several IRAS galaxies in this direction and found that there is a concentration of galaxies
at $cz\sim 13000$ km s$^{-1}$ (they call this concentration as the Monoceros supercluster).
Although the direction of this supercluster is coincident with the position of
MLO concentration C2, we think that they are not physically associated by several reasons described below.  

Histogram of redshifts of the galaxies toward Monoceros (figure 3)
indicates that the galaxies in this direction are separated into three major groups:
the near one with $cz=2200$ -- 2400 km s$^{-1}$, the middle group with  $cz=10000$ -- 12000 km s$^{-1}$,
and  the far group with $cz=18000$ -- 20000 km s$^{-1}$. These three are different clusters of galaxies;
the nearest one is a cluster of galaxies in the Local Supercluster, the second is the Monoceros supercluster \citep{yam93},
and the third is, we call, the  Monoceros NE cluster, because the center of this cluster with  $\sim 18000$ km s$^{-1}$
[around ($\alpha$, $\delta$) =(115$^{\circ}$, $-5^{\circ}$)] is shifted north-east to the center of MLO concentration C2 (see figure 2).
It is possible that the MLO concentration toward Monoceros originates from one of these clusters,
or  from several clusters by a chance coincidence along the line of sight. 

A clue to solve the origin of this MLO concentration is an inner ring 
of 5 MLOs shown by the small ellipse in figure 2. 
At the center of this ellipse, the HII galaxy NGC 2377 is located (indicated by the arrow).  
This is a well studied spiral galaxy (Sbc) with $cz=2449$ km s$^{-1}$, 
and one of strong low-frequency radio sources (3C 178; e.g., \cite{has80}).
There are also several large angular-size galaxies near NGC 2377, though their redshifts are not known
(except LEDA 77124 and 2MASX J07263371$-$0913541; the latter was detected by HI 21cm).
The average angular separation of the 5 MLOs to the ellipse center is approximately 1.5 degree.

We compare this value with the angular separation of a proto-typical radio relic in A 521 \citep{gia06}
and those in RXC J1314.4$-$2515 \citep{ven07}.  They are detected with GMRT at 610 MHz 
by $\sim 23$ mJy beam$^{-1}$ with $HPBW=15''\times 13''$ for A 521 ($z=0.247$), and
by $\sim 16$ mJy beam$^{-1}$ with $HPBW=15''\times 13''$ for RXC J1314.4$-$2515 ($z=0.2474$).    
Though the relic sources for these two are not listed in the VLSS catalog, the VLSS contour maps at 74 MHz 
apparently show enhanced emission of  about 0.5 Jy beam$^{-1}$ at the positions of these relics. 
The angular separation of these relics to the cluster center (approximately 5$'$) can be scaled with redshift as,
\begin{equation}
{\theta _s} \approx 5' (0.247/z ) = 1^{\circ}\ [6200\ {\rm km  \ s}^{-1}/(cz )] , 
\end{equation}
where $z\ll 1$.
The separation of 1.5$^{\circ}$  of the Monoceros ring near NGC 2377
is  comparable and consistent with the scaled separation, 2.6$^{\circ}$, 
of these prototypical radio relics  at $cz\sim 2400$ km s$^{-1}$, if we allow
 the-line-of-sight orientation effect of the separation by about 50\%.

Similarly, we can scale the flux density of the typical relics at the distance of the 2400 km s$^{-1}$ cluster.
If we assume that they are point-like at $z=0.247$, the total flux density, if it is placed 
at the distance of $cz=2400$ km s$^{-1}$, should be about 250 Jy at 74 MHz,
which is much larger than the sum ($\sim 6$ Jy) of peak flux densities 
of  the ringed 5 MLOs towards NGC2377;  even if we include all 20 MLOs of the concentration C2, 
it only goes up to 23 Jy,
though the exact evaluation of the total flux density is difficult because they are not really point-like. 
However, because the low-frequency emission of relics is deduced to be very extended (in a few degree) 
and diffuse at such a small distance, 
it is possible to interpret that  the giant relic was fully resolved and most of the 74 MHz flux was
missing in the VLSS interferometric observations.
Only a few bright patchy portions of filaments are possibly detectable in the VLSS observations\footnote{
In table 2, the brightest source, 0723.4-0904 (2.36 Jy), was resolved in VLSS, 
but not for the other sources. This is probably because  of a low brightness of the extended emission.}.  
In fact,  very filamentary,  giant ringlike double relics without radio halo 
have been found in Abell 3667 ($z=0.055$) and 3376 ($z=0.046$) 
with ATCA or GMRT, respectively,  at 1.4 GHz \citep{rot97,bag06}.  
Unfortunately, no 74 MHz images are available for these objects because of their low declinations
($\delta<-39^{\circ}$). 

\begin{figure}
\begin{center}
\FigureFile(80mm,50mm){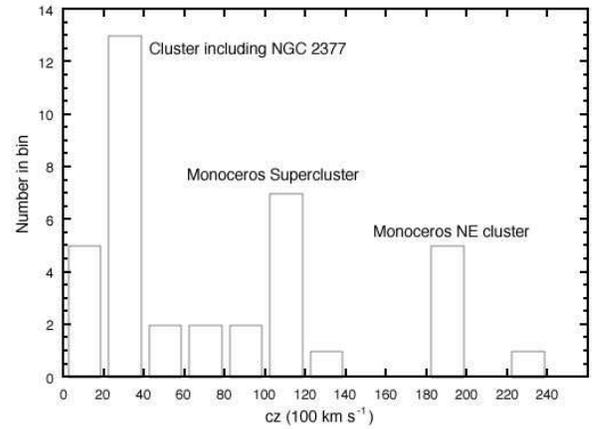}
\end{center}
\caption{Histogram of the radial velocity (cz) for the galaxies with known redshift
toward Monoceros concentration.  
All the galaxies with known redshifts are counted
in a $10^{\circ} \times 10^{\circ}$ area centered at $(l, b)=(225^{\circ}, 4^{\circ})$.
The redshift data were taken from the NASA Extragalactic Database (NED). 
\label{fig:3}
}
\end{figure}

Therefore,  the  MLOs along the inner ellipse in figure 2 are likely to be radio relics associated 
with the galaxy cluster group (at $cz=2400$ km s$^{-1}$) centered on NGC 2377.  
In contrast, the origin of the outer ring of MLOs  (the outer  large  ellipse in figure 2) is much harder to deduce.
It might be an outer shell/ring structure of  the galaxy cluster centered
on NGC 2377. The separations of MLOs  from the center of the ellipse are between  1.7$^{\circ}$ and 4.1$^{\circ}$
(semi-minor and semi-major diameters of the ellipse). Considering that above scaling formula gives a separation of 2.6$^{\circ}$
at the NGC 2377 distance, and allowing for the-line-of-sight orientation effect of the separation,
we can say that they are also a part of radio relics associated with the 2400 km s$^{-1}$ galaxy cluster.
Instead, if they are associated with the Monoceros supercluster, which has $cz\sim 13000$ km s$^{-1}$,
their separation is about 3.5--8.5 times larger than the typical separation between relics and the cluster center.
Therefore, in the condition that all of these clustered MLOs belong to a single physical entity,
 we think that these MLOs are not radio relics in Monoceros supercluster 
 at $cz=13000$ km s$^{-1}$ or beyond.  

However, note that above argument does not exclude the possibility that a few of these MLOs 
are radio relics associated with more distant clusters by chance.  
A  positional coincidence rate of distant MLOs  on the nearby MLO concentration 
can be calculable (see footnote 2).  We computed that 
the number of  background sources does not exceed 7 at most in the 20 MLOs  in this concentration.

There is a small concentration of 4 MLOs around ($\alpha$, $\delta$) =(113.5$^{\circ}$, $-6^{\circ}$)
in figure 2 (the small ellipse in the upper part of figure 2).
However, the redshifts have not been measured for the surrounding galaxies except 2MASX J07322365$-$0548442
at 11063 km s$^{-1}$ \citep{yam93}.  The approximate  separation of MLOs from the center of the second ellipse
is about 0.5$^{\circ}$, which is not inconsistent with the scaled separation of relics to the cluster center if
it is associated to the Monoceros supercluster at $cz\sim 13000$ km s$^{-1}$  .

One of MLOs, 0718.3$-$1120 (the lower right red square near blue circles assigned as 224 in figure 2), 
is located within $15'$ from the X-ray cluster  CIZA J0717.4$-$1119 \citep{ebe02}.
The redshift of this X-ray cluster is  $z=0.0750$ ($cz=22500$ km s$^{-1}$), considerably distant 
from the other two nearby clusters in this direction. 
The separation of 15$'$ from the X-ray cluster center is coincident with the scaled separations
of the relics to the cluster center in  A 521 and  RXC J1314.4$-$2515. Therefore, we cannot exclude the possibility
that VLSS 0718.3$-$1120 is a radio relic associated with the X-ray cluster  CIZA J0717.4$-$1119.

The Monoceros concentration of MLOs  suggests a presence of  a relatively nearby rich galaxy cluster
with a $10^{\circ}$ scale at $cz\sim 2400$ km s$^{-1}$ in this direction, 
though it does not  reveal clearly in figure 2.
Because it is close to the Galactic plane, the past searches for galaxies and measurements of redshifts 
are quite limited in this region.
\citet{vis96} investigated the distribution of galaxies behind the Milkyway in the area,  
$210^{\circ}<l<360^{\circ}$ and $|b|<15^{\circ}$. 
Their figure 4 shows that the dense concentration of galaxies 
with velocities between 1700 and 3200 km s$^{-1}$ is extended at $l=235$ -- 245$^{\circ}$ 
in the the $b<0$ side, but not much  between $l=220$ -- 230$^{\circ}$  in the $b>0$ side,
where the Monoceros MLO concentration (C2) was found.
It is possible that the hypothetical galaxy cluster centered on NGC 2377 at $b=4^{\circ}$ 
is a part of above mentioned large cluster extended in the $b<0$ side,
or it is a cluster colliding and merging with this large cluster.
In summary,  the MLO concentration can be used as a tracer of relatively nearby galaxy clusters.

\section{Conclusion}
We investigated the distribution of the meter-wave luminous objects in the sky,
and found concentrations of these objects in several directions.
In particular, the largest concentration of 20 objects toward Monoceros is used as a clue 
to clarify the origin of the clustering. We conclude that these are 
a part of filamentary structure of giant radio relics
associated with the galaxy cluster at $cz=2400$ km s$^{-1}$,
although a few sources might be  radio relics contaminated from the distant background 
clusters. This fact implies that MLOs detectable at the current sensitivity of
VLA at 74 MHz are useful to map the structure of the Local Supercluster and it's vicinity.   
 
The authors thank  Drs. T. Yamada and H. Nagai for helpful comments. 
This research made use of the SIMBAD and VISIER databases operated at CDS, Strasbourg, France,
 as well as data products from NASA Extragalactic Database (NED) at JPL
 funded by the National Aeronautics and Space Administration and the National Science Foundation. 


\end{document}